\documentclass[11pt]{book}

\usepackage[dvips]{epsfig}
\usepackage{amssymb}
\usepackage{multirow}
\usepackage{amsbsy}
\usepackage{bm}
\usepackage{pstricks}
\catcode`\@=11

 \def\@normalsize{\@setsize\normalsize{13pt}\xipt\@xipt
   \abovedisplayskip 11pt plus3pt minus6pt
   \belowdisplayskip \abovedisplayskip
   \abovedisplayshortskip \z@ plus3pt
   \belowdisplayshortskip 6.6pt plus3.5pt minus3pt}

 \def\small{\@setsize\small{12pt}\xipt\@xipt
   \abovedisplayskip 10pt plus2pt minus5pt
   \belowdisplayskip \abovedisplayskip
   \abovedisplayshortskip \z@ plus3pt
   \belowdisplayshortskip 6pt plus3pt minus3pt
   \def\@listi{\topsep 6pt plus 2pt minus 2pt
     \parsep 3pt plus 2pt minus 1pt
     \itemsep \parsep}}

 \def\footnotesize{\@setsize\footnotesize{10pt}\ixpt\@ixpt
   \abovedisplayskip 8pt plus 2pt minus 4pt
   \belowdisplayskip \abovedisplayskip
   \abovedisplayshortskip \z@ plus 1pt
   \belowdisplayshortskip 4pt plus 2pt minus 2pt
   \def\@listi{\topsep 4pt plus 2pt minus 2pt
      \parsep 2pt plus 1pt minus 1pt
      \itemsep \parsep}}

 \def\scriptsize{\@setsize\scriptsize{9.5pt}\viiipt\@viiipt}
 \def\tiny{\@setsize\tiny{7pt}\vipt\@vipt}
 \def\large{\@setsize\large{14pt}\xiipt\@xiipt}
 \def\Large{\@setsize\Large{18pt}\xivpt\@xivpt}
 \def\LARGE{\@setsize\LARGE{22pt}\xviipt\@xviipt}
 \def\huge{\@setsize\huge{25pt}\xxpt\@xxpt}
 \def\Huge{\@setsize\Huge{30pt}\xxvpt\@xxvpt}
\def\section{\@startsection {section}{1}{\z@}%
{-1.5\baselineskip plus-1pt minus-3pt}{1\baselineskip plus1pt minus2pt}%
{\centering\normalsize\bf}}
\def\subsection{\@startsection{subsection}{2}{\z@}%
{-1\baselineskip plus-1pt minus-2pt}{1\baselineskip plus1pt minus2pt}%
{\normalsize\sc\noindent}}
\def\subsubsection{\@startsection{subsubsection}{3}{\z@}%
{-1\baselineskip plus-1pt minus-2pt}{1sp}{\normalsize\it\noindent}}
\def\paragraph{\@startsection{paragraph}{4}{\z@}%
{1\baselineskip plus1pt minus2pt}{-1em}{\normalsize\it\noindent}}
\let\subparagraph=\paragraph

\def\tableofcontents{\@restonecolfalse\if@twocolumn\@restonecoltrue
\onecolumn\fi\OSIDcont\@starttoc{con}\if@restonecol\twocolumn\fi}

\def\l@section{\@dottedtocline{1}{0em}{.66em}}

\def\thebibliography#1{\section*{{Bibliography}\@mkboth
 {BIBLIOGRAPHY}{BIBLIOGRAPHY}}\footnotesize\rm\list
 {[\arabic{enumi}]}{\settowidth\labelwidth{[#1]}\leftmargin\labelwidth
 \advance\leftmargin\labelsep\usecounter{enumi}}
 \def\newblock{\hskip .11em plus .33em minus -.07em}
 \sloppy\clubpenalty4000\widowpenalty4000
 \sfcode`\.=1000\relax}

\def\ps@myheadings{\let\@mkboth\@gobbletwo
\def\@oddhead{\hbox{}\hfil{\footnotesize\rm\rightmark}\hfil
\normalsize\rm\thepage}\def\@oddfoot{}\def\@evenhead{\normalsize\rm
\thepage\hfil{\footnotesize\rm\leftmark}\hbox{}\hfil
}\def\@evenfoot{}\def\sectionmark##1{}\def\subsectionmark##1{}}

\def\@copyrighthead{}

\def\Year{2008}        %
\newcounter{paPer}     %
\def\EndpagE{\expandafter\pageref{\the\value{paPer}OpSy}}

\def\ps@osiD{\let\@mkboth\@gobbletwo
\def\@oddhead{\@copyrighthead}
  \def\@oddfoot{}\def\@evenhead{}\let\@evenfoot\@oddfoot}

\def\cite{\@ifnextchar [{\@tempswatrue\@Rcitex}{\@tempswafalse\@Rcitex[]}}

\def\@Rcitex[#1]#2{\if@filesw\immediate\write\@auxout{\string\citation{#2}}\fi
  \def\@citea{}\@cite{\@for\@citeb:=#2\do
    {\@citea\def\@citea{,\penalty\@m\,}\@ifundefined
       {b@\the\value{paPer}R\@citeb}{{\bf ?}\@warning
       {Citation `\@citeb' on page \thepage \space undefined}}%
\hbox{\csname b@\the\value{paPer}R\@citeb\endcsname}}}{#1}}

\long\def\@caption#1[#2]#3{\par\addcontentsline{\csname
  ext@#1\endcsname}{#1}{\protect\numberline{\csname
  the#1\endcsname}{\ignorespaces #2}}\begingroup
    \@parboxrestore
    \small                                        %    \normalsize
    \@makecaption{\csname fnum@#1\endcsname}{\ignorespaces #3}\par
  \endgroup}

\newtoks\@stequation

\def\subequations{\refstepcounter{equation}%
\edef\@savedequation{\the\c@equation}%
\@stequation=\expandafter{\theequation}%   %only want \theequation
\edef\@savedtheequation{\the\@stequation}% %expanded once
\edef\oldtheequation{\theequation}%
\setcounter{equation}{0}%
\def\theequation{\oldtheequation\alph{equation}}}%

\def\endsubequations{%
\setcounter{equation}{\@savedequation}%
\@stequation=\expandafter{\@savedtheequation}%
\edef\theequation{\the\@stequation}\global\@ignoretrue}

\catcode`\@=12
\let\Rlabel=\label
\let\Rbibitem=\bibitem
\let\Rref=\ref
\let\Rpageref=\pageref
\def\label#1{\expandafter\Rlabel{\the\value{paPer}R#1}}
\def\bibitem#1{\expandafter\Rbibitem{\the\value{paPer}R#1}}
\def\ref#1{\expandafter\Rref{\the\value{paPer}R#1}}
\def\pageref#1{\expandafter\Rpageref{\the\value{paPer}R#1}}

\def\thesection{\arabic{section}.}

\def\YYMm{\rule{0ex}{4em}}
\newtoks\TITsi
\newtoks\TITsii

\def\title#1{\def\TITs{\LARGE{\raggedright\noindent\YYMm #1%
\vskip8pt\par}}}
\def\author#1{\autMM{#1}\def\LHD{#1}}
\def\and{{\rm\lowercase{and}}}

\def\autMM#1{\TITsii={\vskip10pt\par\normalsize\rm\noindent #1\par}%
\TITsi=\expandafter{\TITs}\edef\TITs{\the\TITsi\the\TITsii}}

\def\address#1{\TITsii={\vskip6pt\par\footnotesize\sl\noindent #1\par}%
\TITsi=\expandafter{\TITs}%
\edef\TITs{\the\TITsi\the\TITsii}}

\def\received#1{\TITsii={\vskip10pt\par\small\rm\noindent(Received: #1)\par}%
\TITsi=\expandafter{\TITs}\edef\TITs{\the\TITsi\the\TITsii}}

\def\headtitle#1{\def\RHD{#1}}
\def\headauthor#1{\def\LHD{#1}}
\def\listas#1#2{\addcontentsline{con}{section}{{\sc #1: }{\rm #2}}}

\def\abst{{\bf Abstract.}}
\def\abstract#1{\TITs
       \vskip15pt\par\noindent
       {\footnotesize{\abst~} #1\vskip3pt\par}
       \markright{\RHD}
       \markboth{\LHD}{\RHD}}

\def\startpaper{%
       \cleardoublepage
       \setcounter{section}{0}
       \stepcounter{paPer}
       \setcounter{equation}{0}
       \setcounter{footnote}{0}
       \setcounter{figure}{0}
       \setcounter{table}{0}
       \def\theequation{\arabic{equation}}
       \def\thefootnote{\arabic{footnote}}
       \setcounter{defn}{0}
       \setcounter{thm}{0}
       \setcounter{lem}{0}
       \setcounter{prop}{0}
       \setcounter{rem}{0}
       \thispagestyle{osiD}}

\def\OSIDcont{\cleardoublepage\thispagestyle{empty}
       \markright{}\markboth{}{}
       \normalsize\rm
       \hspace*{\fill}{\large\rm
         Contents of the Volume \Volume, Number \Number}\hspace*{\fill}
       \par\vspace{1.5em}
       \par\noindent}

\def\endpaper{\expandafter\label{\the\value{paPer}OpSy}}

\def\1{{\mathchoice{\rm 1\mskip-4mu l}{\rm 1\mskip-4mu l}%
{\rm 1\mskip-4.5mu l}{\rm 1\mskip-5mu l}}}
\def\varkappa{\mbox{\bBB\char 123}}
\def\longhookrightarrow{\lhook\joinrel\relbar\joinrel\rightarrow}

\def\longhookUp{\lower6pt\hbox{\rotatebox{90}{$\longhookrightarrow$}}}

\def\Tr{\mathop{\rm Tr}}

\def\theequation{\thesection\arabic{equation}}

\def\Myskip{\setlength{\baselineskip}{13pt}}
\def\text#1{\quad\mbox{\rm  #1 }\quad}

\input xy
\xyoption{all}

\InputIfFileExists{psfig.sty}{\typeout{^^Jpsfig.sty inputed...ok}}{\typeout{^^JWarning: psfig.sty could not be found.^^J}}
\begin{document}

\def\artid{0000001}
\def\Volume{1}
\def\Number{1}
\def\Year{2017}

\startpaper

\definecolor{dgreen}{rgb}{0,0.5,0}
\definecolor{delete}{cmyk}{0.5,0,0,0}
\newcommand{\RED}[1]{{\color{red}#1}}
\newcommand{\BLUE}[1]{{\color{blue}#1}}
\newcommand{\REV}[1]{{\color{red}#1}}
\newcommand{\GREEN}[1]{{\color{dgreen}#1}}
\newcommand{\DEL}[1]{{\color{delete}#1}}
\newcommand{\KY}[1]{{\color{dgreen}\textbf{[[[#1]]]}}}

%%%%%

%\def\cH{{\mathcal{H}}}
%\def\cV{{\mathcal{V}}}
%\def\cU{{\mathcal{U}}}
%\def\cP{{\mathcal{P}}}
%\def\cN{{\mathcal{N}}}
%\def\cS{{\mathcal{S}}}
%\def\cR{{\cal R}}
%\def\cZ{{\cal Z}}
%\newcommand{\As}{\mathcal{A}}
%\newcommand{\bm}[1]{\mbox{\boldmath $#1$}}

%\def\As{{\mathcal{A}}}
%\def\bra#1{\langle #1 |}
%\def\ket#1{| #1 \rangle}
%\def\bbra#1{( #1 |}
%\def\kket#1{| #1 )}
%\def\Ord{\mathrm{O}}
\renewcommand{\Re}{\mathop{\mathrm{Re}}}
\renewcommand{\Im}{\mathop{\mathrm{Im}}}

%\newcommand{\REV}[1]{\textbf{\color{red}#1}}
%\newcommand{\BLUE}[1]{\textbf{\color{blue}#1}}
%\newcommand{\GREEN}[1]{\textbf{\color{green}#1}}
%\newcommand{\VIOLET}[1]{\textbf{\color{violet}#1}}

%%%%%

\newcommand{\beq}{\begin{equation}}
\newcommand{\eeq}{\end{equation}}
\newcommand{\barr}{\begin{eqnarray}}
\newcommand{\earr}{\end{eqnarray}}

\def\e{\mathrm{e}}
\def\ii{\mathrm{i}}
\def\d{\mathrm{d}}

\title{Can decay be ascribed to classical noise?}
\author{Daniel Burgarth$^1$, Paolo Facchi$^{2,3}$, Giancarlo Garnero$^{2,3}$, Hiromichi Nakazato$^{4}$, Saverio Pascazio$^{2,3}$ and Kazuya Yuasa$^{4}$}
%email: burgarth@gmail.com
%email:  paolo.facchi@ba.infn.it 
%email:  giancarlo.garnero@ba.infn.it
%email:  hiromici@waseda.jp}
%email:  saverio.pascazio@ba.infn.it

\address{$^1$ Institute of Mathematics, Physics and Computer Science, Aberystwyth University,\\
Aberystwyth SY23 3BZ, UK}
\address{$^2$ Dipartimento di Fisica and MECENAS, Universit\`a di Bari, I-70126 Bari, Italy} 
\address{$^3$ INFN, Sezione di Bari, I-70126 Bari, Italy}
\address{$^4$ Department of Physics, Waseda University, Tokyo 169-8555, Japan} 

\headauthor{D. Burgarth, P. Facchi, G. Garnero, H. Nakazato, S. Pascazio and K. Yuasa}
\headtitle{Can decay be ascribed to classical noise?}
\received{\today}
\listas{D. Burgarth, P. Facchi, G. Garnero, H. Nakazato, S. Pascazio and K. Yuasa}{Self-dual vs non-self-dual maps}

\abstract{No.}

\Myskip

%\pacs{03.65.Yz}	%Decoherence; open systems; quantum statistical methods,
%42.50.Lc Quantum fluctuations, quantum noise, and quantum jumps}

\section{Introduction and motivation}
The dynamics of a dissipative quantum system, in the Markovian approximation, is governed by the Gorini-Kossakowski-Lindblad-Sudarshan (GKLS) equation~\cite{Kossa1972,GKS,Lind}. This equation always admits a dilation to a stochastic differential equation \cite{HP,AF,KM87} 
and can be read as a (quantum) Langevin equation~\cite{GZ}. In the mathematical physics literature, stochastic equations have been studied both for Gaussian processes
\cite{AF} and for general combinations of Gaussian and Poisson processes \cite{KM87}. In this article, we shall limit our analysis to the Gaussian case, which is more relevant for physical applications, see for example the review~\cite{Gough} on derivations and applications of stochastic Schr\"odinger equations for quantum control and quantum information processing.

Quantum dissipation can take different forms, and is associated with different physical scenarios. Among these, there are genuine ``dephasing'' processes, as well as \emph{bona-fide} ``decay'' processes (e.g., to the ground state). Accordingly, the GKLS equations have different mathematical features and physical meaning: 
for instance, some physical features of dephasing are often reflected in the self-duality of the quantum dynamical map.

We ask here the following questions: are these different physical and mathematical features mirrored in the Wiener process associated with the corresponding quantum Langevin equation? More specifically: can decay be ascribed to ``classical'' noise? Moreover: do the afore-mentioned features affect the Hamiltonians of the associated Ito and Stratonovich stochastic equations, and if yes, how? The answers to the above questions will require definition and physical elucidation of these concepts. On this basis, we will endeavour to clarify the physical meaning of the corresponding stochastic Schr\"odinger equations.

This article is organized as follows.
In Sec.~\ref{sec-gendef}\ we introduce notation and review paradigmatic examples of self-dual (SD) and non-self-dual (NSD) maps. We start with a few observations in Sec.~\ref{sec:theorem}, where we give a physical definition of classical noise. In Sec.~\ref{sec:more} we look at a rather general example, that involves both SD and NSD components. We prove our first main result in Sec.\ \ref{sec:gentheorem}, where the Stratonovich formulation is also discussed.
The relation between self-duality and dephasing is elaborated in Sec.\ \ref{sec:new}, where a definition is given of  dephasing and decay channels and general conclusions are drawn.
We put forward a few additional remarks in Sec.~\ref{sec:correlated}
and answer the question posed in the title in Sec.~\ref{sec-concl}.

\section{Generalities and definitions}
\label{sec-gendef}

The GKLS evolution equation for the density matrix $\varrho$ of a quantum system reads
\begin{equation}
\label{Msol}
\dot \varrho(t) = \mathcal{L}_\mathrm{tot} \varrho(t)  , \qquad \mathcal{L}_\mathrm{tot} = \mathcal{L}_H + \mathcal{L},
\end{equation}
where $\mathcal{L}_H$ and  $\mathcal{L}$ are the (time-independent) Hamitonian and dissipative parts of the total map $\mathcal{L}_\mathrm{tot}$, respectively, 
and the dot denotes derivative with respect to time $\mathrm{d}/\mathrm{d} t$.
The solution is
\begin{equation}
\label{Msol2}
\varrho(t)  = \e^{t \mathcal{L}_\mathrm{tot}} \varrho(0) = \Lambda_{t} \varrho(0) \quad (t \geqslant 0).
\end{equation}
The adjoint dynamical equation for an observable $A$ is given by
\begin{equation}
\label{M2}
    \dot{A}(t) = \mathcal{L}_\mathrm{tot}^{\sharp} A(t)  ,
\end{equation}
whose solution is
\begin{equation}
\label{M3}
 A(t) =  \Lambda_{t}^\sharp A(0) \quad (t \geqslant 0).
\end{equation}
The Dirac prescription \cite{Diracbook}
\begin{equation}
\label{SvsH1}
\Tr[\varrho(t) A(0)] = \Tr[\varrho(0) A(t)],\quad\forall \varrho,A
\end{equation}
connects the Schr\"odinger and Heisenberg pictures and consecrates their equivalence.
The dissipative component $\mathcal{L}$ of the map is said to be self-dual if 
\begin{equation}
\label{SvsH}
\mathcal{L} = \mathcal{L}^{\sharp} \quad \Longleftrightarrow \quad 
\Lambda_{t}=\Lambda^\sharp_{t} ,
\end{equation}
while it is non-self-dual otherwise.

\subsection{Example of self-dual map: phase damping}
\label{sec-sd}
Typical examples of self-dual maps are those describing phase damping. Consider for example the phase damping of a qubit performing Rabi oscillations. The evolution of the density matrix of the qubit is described by $(\hbar=1)$
\begin{equation}
\dot \varrho = \mathcal{L}_\mathrm{tot}\varrho = - \ii [\Omega \sigma_1, \varrho] -  \gamma (\varrho- \sigma_3 \varrho \sigma_3),
\label{qubit3}
\end{equation}
where $\gamma>0$, and $\sigma_\alpha \, (\alpha = 0,1,2,3)$ are the Pauli matrices (with $\sigma_0\equiv \mathbf{1}$). The adjoint equation for an observable $A$ reads
\begin{equation}
\label{M22}
 \dot{A} = \mathcal{L}_\mathrm{tot}^{\sharp} A  = \ii [\Omega \sigma_1, A] - \gamma (A- \sigma_3 A \sigma_3)
\end{equation}
and its dissipative part is SD, as $\mathcal{L} = \mathcal{L}^{\sharp}$. 
Physically, the above equation describes Rabi oscillations accompanied by a dephasing process: for example, if $\Omega=0$, the asymptotic solution of Eq.~(\ref{qubit3}) reads
\begin{equation}
\label{qubit3sol}
   \varrho = \frac{1}{2} (\sigma_0 + \bm x \cdot \bm\sigma) \stackrel{t \to \infty}{\longrightarrow}  
   \varrho(\infty) = \frac{1}{2} (\sigma_0 + x_3 \sigma_3),
\end{equation}
$\bm x = \Tr (\rho\bm\sigma)$ being the 3-dimensional Bloch  vector, $| \bm x | \leqslant 1$. When $\Omega=0$ populations do not change, but dephasing makes interference (between eigenstates of $\sigma_3$) impossible.

Equation~(\ref{qubit3}) can be derived from the stochastic Hamiltonian \cite{NP99}
\beq
H_\eta = \Omega \sigma_1 + \sqrt{\gamma} \,\eta (t) \sigma_3,
\label{eq:BBHam}
\eeq
where $\eta$ is a white noise (precise definitions are given later). 
The corresponding stochastic Schr\"odinger equation reads, by Ito calculus,
\barr
\d\psi  &=& -\ii \Omega \sigma_1 \psi \, \d t
                  -\ii \sqrt{\gamma} \, \sigma_3  \psi  \circ \d W \nonumber \\
                &=& [ -\ii \Omega \sigma_1 - (\gamma/2) ]
\psi \, \d t -\ii \sqrt{\gamma} \, \sigma_3  \psi  \,\d W,
\label{eq:strat1}
\earr
where $\circ$ denotes the Stratonovich  product and $W = \int \eta \,\d t$ is the Wiener process. 
Equation~(\ref{eq:strat1}) yields Eq.~(\ref{qubit3}) on average over the realizations of the Wiener process.

A similar example is the phase damping of a harmonic oscillator, whose dissipative part reads
\begin{equation}
\label{oscphdampnew}
   \mathcal{L} \varrho =  -\frac{\gamma}{2}(\{N^2,\varrho \} -2 N\varrho N ),
\end{equation}
where $N=a^\dagger a$ and $[a,a^\dagger]=1$. Again, $\mathcal{L}^{\sharp}=\mathcal{L}$ and the map is SD\@.
If the Hamiltonian is $H = \Omega a^\dagger a$, a generic density matrix becomes diagonal in the $N$-representation
\begin{equation}
\label{rho0}
\varrho =  \sum c_{mn}  |m \rangle \langle n|  \, \stackrel{t \to \infty}{\longrightarrow} \, \varrho(\infty) = \sum c_{nn}  |n \rangle \langle n| ,
\end{equation}
so that populations do not change, but interference among eigenstates of the number operator becomes impossible.

%%%%%%%%%%%%%%%%%%%%%%%%%%%%%%%%%%%%%%%%%%%%%%%%%%%%%%%%%%%%%%%%
\subsection{Example of non-self-dual map: energy damping}
\label{sec-nsd}
Typical examples of non-self-dual maps are those describing energy damping. Consider for example the energy damping of a qubit. 
Let
\begin{equation}
\dot \varrho = \mathcal{L}_\mathrm{tot} \varrho = - \ii [H, \varrho]  -\frac{\gamma'}{2}(\{\sigma_+ \sigma_-,\varrho\}-2\sigma_-\varrho\sigma_+),
\label{Ldec}
\end{equation}
where $\sigma_\pm=(\sigma_1 \pm\ii \sigma_2)/2$ and $H$ is a Hamiltonian. One has 
\begin{equation}
\mathcal{L}_\mathrm{tot}^{\sharp} A= \ii [H, \varrho]  - \frac{\gamma'}{2}(\{\sigma_+ \sigma_-,A\}-2\sigma_+A\sigma_-),
\end{equation}
and the dissipative evolution is non-self-dual: $\mathcal{L}^{\sharp} \neq \mathcal{L}$.
If $H = (\Omega/2) \sigma_3$, the solution of Eq.~(\ref{Ldec}) reads
\begin{equation}
\label{qubit3solbis}
   \varrho = \frac{1}{2} (\sigma_0 + \bm x \cdot \bm\sigma)  \stackrel{t \to \infty}{\longrightarrow} 
   \varrho(\infty)= P_-,
\end{equation}
so that the final state is the projection $P_- = (\sigma_0 - \sigma_3)/2$ over the ground state. 
Equation~(\ref{Ldec}) must be derivable \cite{HP,AF} from a stochastic (non-Hermitian) Hamiltonian, through a term of the type 
\beq
H_\eta = H + \ii \sqrt{\gamma'} \, \eta (t) \sigma_- .
\label{eq:decHam}
\eeq
However, such a derivation is not conceptually painless, as we shall see in the following.

A related example is the energy damping of a harmonic oscillator. For this dynamics, we have
\begin{equation}
\label{oscdamp}
\mathcal{L}\varrho =  -\frac{\gamma'}{2}( \{a^\dagger a,\varrho \} -2 a\varrho a^\dagger),
\end{equation}
whence
\begin{equation}
\mathcal{L}^{\sharp} A =  -\frac{\gamma'}{2} ( \{a^\dagger a,A \} -2 a^\dagger A a ),
\end{equation}
which is NSD\@. The oscillator decays to the ground state (e.g., for $H=\Omega a^\dag a$)
\begin{equation}
\label{rho1}
\varrho =  \sum c_{mn}  |m \rangle \langle n|  \stackrel{t \to \infty}{\longrightarrow}  \varrho(\infty) =  |0 \rangle \langle 0| .
\end{equation}
One of the main objectives of the present article is to elucidate whether the difference between dephasing processes (and in general SD maps of the type shown in Sec.\
\ref{sec-sd}), and decay processes (and in general NSD maps of the type shown in the present section) are reflected in some structural properties of the associated stochastic Schr\"odinger equation. 

Incidentally, we observe that the infinite-time limit of the dissipative dynamics may lead to a contraction of the algebra of observables \cite{CFMP,ACFMPR}. 
We will not discuss in this article whether such a contraction may bear signatures of the self-duality of the map (or lack thereof).

%%%%%%%%%%%%%%%%%%%%%%%%%%%%%%%%%%%%%%%%%%%%%%%%%%%%%%%%%%%%%%%%
\section{A few observations}
\label{sec:theorem}

Let $\eta$ be a white noise 
\beq
\langle \eta (t) \rangle = 0, \qquad
\langle \eta (t) \eta(t') \rangle = \delta (t-t'),
\label{eq:wn}
\eeq
the brackets denoting the ensemble average over all possible realizations of the noise. The associated Wiener process reads
\begin{equation}
\d W(t) \equiv W(t+\d t)-W(t) = \int_t^{t+\d t} \eta(s)\d s ,
\end{equation}
\begin{equation}
\langle \d W(t) \rangle = 0, \quad
\langle \d W(t) \d W(t) \rangle = \d t.
\label{eq:wiep}
\end{equation}
Consider the stochastic Schr\"odinger equation
\begin{equation}
\d\psi=-\ii (H_1 -\ii H_2) \psi \,\d t
                     -\ii L \psi \,\d W 
\label{eq:strat}
\end{equation}
to be understood in the Ito sense. The operators $H_1$ and $H_2$ are taken to be Hermitian while, in general, $L$ is not.

The evolution of the density matrix  $\rho =  |\psi \rangle \langle \psi|$ is governed by
\barr
\d\rho &=& |\psi +\d\psi \rangle \langle \psi+ \d\psi| -
          |\psi \rangle \langle \psi| \nonumber \\
   & = &  -\ii[H_1,\rho]\d t - \{H_2,\rho \}\d t
          -\ii(L\rho-\rho L^\dagger)\d W + L\rho L^\dagger \d t,
\label{eq:evrho}
\earr
where $[\,\cdot\,,\,\cdot\,]$ is the commutator and  $\{\,\cdot\,,\,\cdot\, \}$
the anticommutator.

By taking the trace we get
\begin{eqnarray}
\Tr (\d\rho) &=&  \| \psi+\d\psi \|^2- \|\psi \|^2 
\nonumber\\
&=& \langle \psi | (-2 H_2 + L^\dagger L) \psi\rangle  \d t  -\ii  \langle \psi | (L- L^\dagger) \psi\rangle \d W, 
\end{eqnarray}
and, by taking the average over the noise, we see that a ``weak" (i.e.\ on average)
conservation of probability 
\beq
\langle \| \psi+\d\psi \|^2 \rangle =
\langle \|\psi \|^2 \rangle =1
\label{eq:probOK}
\eeq
imposes a definite relation between the noise term and the non-Hermitian part of the Hamiltonian: 
\beq
H_2= \frac{1}{2}L^\dagger L.
\label{eq:H2}
\eeq
This yields, by taking the average of~(\ref{eq:evrho}), a GKLS equation 
\beq
\frac{\d}{\d t} \rho=
          -\ii[H_1,\rho] - \frac{1}{2}
                    \{ L^\dagger L,\rho \} +
                    L \rho L^\dagger .
\label{eq:Lindy}
\eeq

Notice that, while  relation~(\ref{eq:H2}) implies that probability is conserved on average, in general probability is \emph{not conserved} along each \emph{single realization} of the noise, which, therefore, does not represents a physical evolution. 
Every trajectory is physical and probability is (strictly) conserved in each individual realization if and only if 
\begin{equation}
L=L^\dagger,
\label{eq:LLdag}
\end{equation}
which, in turn, implies that the dissipative part of the generator is self-dual
\begin{equation}
\mathcal{L}=\mathcal{L}^\sharp.
\end{equation}
When this happens, one can describe the dynamics in terms of a 
(Hermitian) time-dependent Hamiltonian
\beq
H_\eta = H_1 + \eta (t) L,
\label{eq:iHam}
\eeq
where $H_1$ and $L$ are Hermitian, time-independent operators.
Observe also that in this case the stochastic Schr\"odinger equation~(\ref{eq:strat}) admits a natural expression in terms of the 
Stratonovich product $\circ$ 
\begin{equation}
\d\psi=-\ii H_1  \psi \,\d t
                     -\ii L \psi \circ \d W .
\label{eq:strat1L}
\end{equation}
Equations~(\ref{eq:LLdag})--(\ref{eq:strat1L}) enable one to speak of a ``classical" noise, in the sense that one can view the dynamics as arising from a classical noise source. An example will elucidate the concept. Consider a spin-1/2 particle in a magnetic field $\bm B$, described by the Hamiltonian 
\beq
H =  {\bm \mu} \cdot {\bm B} ,
\label{eq:HamB}
\eeq
where $\bm \mu = \mu \bm \sigma$ is the magnetic moment. If the magnetic field has a random (white noise) component
\beq
\bm B = \bm B_0 + \bm \delta {\bm B (t)} ,
\label{eq:ranB}
\eeq
then
\beq
H =  \bm \mu \cdot \bm B_0 + \sqrt{\gamma}  \eta (t) \bm \sigma \cdot {\bm n} ,
\label{eq:Hamspell}
\eeq
${\bm n}$ being the unit vector (assumed to be time-independent) parallel to the random component of the field, and $\gamma \propto \delta B$. This Hamiltonian has the form (\ref{eq:iHam}) with $L=L^\dagger$, and describes the effect of a (classical) noisy magnetic field. From a physical perspective, the energy eigenvalues appear to be ``shaken" by a random component.

Notice also that in this case there is no need of taking the average in Eq.~(\ref{eq:probOK}), as 
\beq
 \| \psi+\d\psi \|^2  = \|\psi \|^2  =1
\label{eq:probOK2}
\eeq
in every individual realization of the stochastic process [provided Eq.\ (\ref{eq:H2}) holds].
Physically, one can view the quantum system as governed by a \emph{bona fide} (Hermitian) time-dependent Hamiltonian~(\ref{eq:iHam}) $\forall \eta$. The dynamics is always unitary and probabilities are always conserved.

\section{One additional example: interaction with a thermal field}
\label{sec:more}

Let us look at one additional example: a two-level atom in interaction with a thermal field and subject to dephasing and decay. This example involves both SD and NSD components and puts together examples~(\ref{qubit3}) and~(\ref{Ldec}) of the preceding section, generalizing the latter to non-vanishing temperature.

The dynamics is given by $H_1 = (\Omega/2) \sigma_3$ and
\begin{eqnarray}
   \mathcal{L} \varrho &=& - \frac{\gamma'}{2} (1+n) (  \{\sigma_+\sigma_-, \varrho\} - 2 \sigma_-\varrho \sigma_+)
     \nonumber \\
    & &- \frac{\gamma'}{2} n (\{\sigma_-\sigma_+, \varrho\} - 2 \sigma_+\varrho \sigma_- ) - \gamma (\varrho- \sigma_3 \varrho \sigma_3),
\label{decay2}
\end{eqnarray}
where $n=(\e^{\beta  \Omega} -1)^{-1}$, with $\beta$ the inverse temperature and $\Omega$ the energy difference of the two atomic states, and $\gamma \propto \beta^{-1}$. The constants $\gamma$ and $\gamma^\prime$ are independent \cite{FTPNTL}.

The asymptotic solution of Eq.~(\ref{decay2}) reads
\begin{equation}
\label{qubitthsol}
\varrho = \frac{1}{2} (\sigma_0 + \bm x \cdot \bm\sigma) \stackrel{t \to \infty}{\longrightarrow}  \varrho(\infty) = \frac{P_- + \e^{-\beta  \Omega} P_+}{1+ \e^{-\beta  \Omega}} ,
\end{equation}
where the notation is the same as in Eq.~(\ref{qubit3sol}), $P_\pm = \sigma_\pm \sigma_\mp = (\sigma_0 \pm \sigma_3)/2$ are the two projections, and Boltzmann's statistics is implied.
The stochastic Ito-Schr\"odinger equation reads
\begin{equation}
\d\psi =- i(H_1-iH_2) \psi \,\d t
                    - \ii \left(L_- \d W_- + L_+ \d W_+  + L_3 \d W_3 \right) \psi , 
\label{eq:strat3}
\end{equation}
where the noises are independent, $\langle \d W_k \rangle= 0$, 
$\langle \d W_k \d W_l\rangle=\delta_{kl}\d t$, with $k,l=\pm,3$, and 
\begin{eqnarray}
L_- & = & -\ii \sqrt{ \gamma' (n+1)}\, \sigma_- , \nonumber \\
L_+ & = & -\ii \sqrt{ \gamma' n}\, \sigma_+ , \label{eq:genexpl} \\
L_3 & = & \sqrt{ \gamma}\,\sigma_3 , \nonumber 
\end{eqnarray}
so that (weak) probability conservation implies 
\begin{equation}
H_2 = \frac{1}{2} \sum_{k=\pm,3} L^{\dagger}_k L_k = \frac{\gamma'}{2} (1+n) P_+ + \frac{\gamma'}{2} n P_- + \frac{\gamma}{2} \sigma_0
\label{eq:h2itogenexpl} 
\end{equation}
in agreement with the GKLS equation (\ref{decay2}), as it should.

As in the examples considered in Sec.\ \ref{sec-gendef}, similar comments apply to the thermal damping of a harmonic oscillator (with $H_1=\Omega a^\dag a$ and $N=a^\dagger a$)
\begin{eqnarray}
\label{damp}
   \mathcal{L} \varrho &=& -\frac{\gamma'}{2} (1+n) ( \{a^\dagger a,\varrho \} -2 a\varrho a^\dagger ) 
\nonumber \\
    & &    
     -\frac{\gamma'}{2} n ( \{aa^\dagger,\varrho \} -2 a^\dagger \varrho a) -\frac{\gamma}{2} (\{N^2,\varrho \} -2 N\varrho N ).   
\end{eqnarray}

\section{Generalization and first main theorem}
\label{sec:gentheorem}
We now generalize the observations of Secs.\ \ref{sec:theorem} and \ref{sec:more} to the case of a master equation with $N$ GKLS operators $L_k$ $(k=1,\dots,N$). Notice that it is sufficient to consider $N\leq d^2-1$, where $d$ is the dimension of the Hilbert space. A larger number of operators will be dependent and always reducible to this case.

In the Ito form, the stochastic Schr\"odinger equation reads
\begin{equation}
\d\psi=-\ii (H_1-\ii H_2)\psi\,\d t-\ii\sum_{k=1}^N L_k \psi \,\d W_k,
\label{eqn:SSEL}
\end{equation}
where $H_1=H_1^\dagger$, $H_2=H_2^\dagger$, and, in general,  $L_k\ne L_k^\dagger$. Moreover, the noises are taken to be normalized and independent:
\begin{equation}
\langle \d W_k \rangle= 0, \qquad 
\langle \d W_k \d W_l\rangle=\delta_{kl}\d t.
\label{eq:noisesgen}
\end{equation}
From the weak conservation of probability~(\ref{eq:probOK}) we get 
\begin{equation}
H_2 = \frac{1}{2}\sum_k L^{\dagger}_k L_k 
\label{eq:h2itogen}
\end{equation}
and the ensuing master equation 
\begin{eqnarray}
\frac{\d\rho}{\d t} &=& - \ii[H_1,\rho]-\{H_2,\rho \}+ \sum_k  L_k\rho L^{\dagger}_k  \nonumber \\
&=& - \ii[H_1,\rho]-\frac{1}{2}\sum_{k}(\{L_k^{\dagger}L_k,\rho\}-2L_k\rho L_k^\dagger).
\label{me}
\end{eqnarray}

The Stratonovich form of the stochastic Schr\"odinger equation is instead
\begin{equation}
\d \psi= -\ii(H_1^S- \ii H_2^S)\psi \,\d t -\ii \sum_{k=1}^N L_k \psi\circ \d W_k ,
\end{equation}
where $H_1^S - \ii H_2^S= H_1-\ii H_2 + \frac{\ii}{2}\sum_k L^2_k$, that is,
\begin{equation}
H_1^S= H_1-\frac{1}{2}\Im\sum_k L^2_k=H_1-\frac{1}{2}\sum_k\frac {L^2_k-{L^2_k}^\dagger}{2\ii}
\label{eq:h1s}
\end{equation}
and
\begin{equation}
H_2^S= H_2-\frac{1}{2}\Re\sum_k L^2_k=\frac{1}{2}\sum_k\left( L^{\dagger}_k L_k-\frac {L^2_k+{L^2_k}^\dagger}{2}\right).
\label{eq:h2s}
\end{equation}
Therefore, the total Hamiltonian reads
\begin{equation}
\label{eq:ham}
H_{\eta} =H_1^S - \ii H_2^S+\sum_k\eta_k(t) L_k.
\end{equation}
This Hamiltonian is Hermitian if and only if
\begin{equation}
L_k=L_k^\dagger, \quad \forall k.
\label{eq:HermLk}
\end{equation}
Indeed, this condition implies  that the non-Hermitian time-independent Hamiltonian vanishes, namely,
\begin{equation}
H_2^S=\frac{1}{2}\sum_k\left( L^{\dagger}_k L_k-\frac {L^2_k+{L^2_k}^\dagger}{2}\right)=0.
\label{eq:HermHS}
\end{equation}
Remarkably, conditions~(\ref{eq:HermLk}) and~(\ref{eq:HermHS}) are in fact \emph{equivalent}, as one can easily prove by setting $L_k=X_k+\ii Y_k$ and taking the trace.
This is an instance of the fluctuation-dissipation theorem: a (non-)Hermitian time-independent Hamiltonian (i.e., an imaginary optical potential \emph{\`a la} Fermi \cite{fermizinn,PTorun}) in Eq.~(\ref{eq:ham})  is accompanied by a (non-)Hermitian noise term.

Conversely, one can derive a \textit{bona fide} GKLS equation from a non-Hermi\-tian dissipative Hamiltonian $H - \ii V$ by adding an anti-Hermitian fluctuating term with  $L= \ii V^{1/2}$. This is a way to cure the illness of an optical potential by restoring probability conservation through a fluctuation-dissipation mechanism.

The dissipative part of the master equation~(\ref{me}) reads
\begin{equation}
\mathcal{L}\rho=-\frac{1}{2}\sum_{k}(\{L_k^{\dagger}L_k,\rho\}-2L_k\rho L_k^\dagger),
\label{eq:diagl}
\end{equation} 
hence its dual is
\begin{equation}
\mathcal{L}^\sharp A = -\frac{1}{2}\sum_{k}(\{L_k^{\dagger}L_k,A\}-2L_k^\dagger A L_k).
\label{eq:diagldual}
\end{equation}
We recall that the trace conservation property, $\Tr(\mathcal{L}\rho)=0$, is equivalent to the unitality of the dual map,  $\Lambda^\sharp \mathbb{I}=\mathbb{I}$ or $\mathcal{L}^\sharp \mathbb{I}=0$.

By looking at the above expressions it is evident that $L_k=L_k^\dagger$ implies the self-duality of $\mathcal{L}$, say $\mathcal{L}=\mathcal{L}^\sharp$. The converse does not hold due to the non-uniqueness of the decomposition of $\mathcal{L}$ in terms of the GKLS operators $L_k$: for example, $\mathcal{L}^\sharp=\mathcal{L}$ when $L_k^\dagger= \e^{\ii \alpha_k}L_k$, with arbitrary phases $\alpha_k$. 
Summarizing, we arrive at the following conclusion:
\begin{equation}
H_\eta=H^\dagger_\eta \iff H_2^S=0 \iff L_k=L_k^\dagger,   \; \forall k
\ \Longrightarrow\ \mathcal{L}=\mathcal{L}^\sharp.
\label{eq:centrth}
\end{equation}

Conditions (\ref{eq:ham})--(\ref{eq:HermHS}), together with (\ref{eq:centrth}), generalize Eqs.~(\ref{eq:LLdag})--(\ref{eq:iHam}). The validity of these conditions enable one to say that the noise sources in Eq.\ (\ref{eq:ham}) are ``classical". 

Let us check the main conclusions of this section by looking again at the example of a two-level atom in interaction with a thermal field, discussed in Sec.\ \ref{sec:more}.
Let us translate this example in Stratonovich form and check the chain of equivalence~(\ref{eq:centrth}). 
In the case (\ref{eq:strat3})--(\ref{eq:h2itogenexpl}), Eqs.\ (\ref{eq:h1s})--(\ref{eq:h2s}) specialize to 
\begin{eqnarray}
& & H_1^S = 0 \quad (\textrm{since} \; H_1 = 0),  \\
& & H_2^S = \frac{\gamma'}{2} (1+n) P_+ + \frac{\gamma'}{2} n P_- ,
\label{eq:HermHS2}
\end{eqnarray}
so that, according to Eq.\ (\ref{eq:ham}),
\begin{equation}
\label{eq:ham22}
H_{\eta} = \ii H_2^S+\sum_{k=\pm,3} \eta_k(t) L_k,
\end{equation}
with $L_k$'s given in Eq.\ (\ref{eq:genexpl}).
Notice that $H_2^S$ does not vanish, due to the presence of the NSD components $L_\pm$, that yield the terms in Eq.\ 
(\ref{eq:HermHS2}). This makes the interpretation of the Stratonovich ``Hamiltonian" cumbersome for NSD equations. 
Incidentally, this example also clarifies that the ``Hamiltonians" (\ref{eq:BBHam}), (\ref{eq:decHam}) and 
(\ref{eq:ham22}) require different physical interpretations.
In general, if condition (\ref{eq:HermLk}) does not hold, then the Stratonovich Hamiltonian (\ref{eq:HermHS}) does not vanish.

The following two sections are devoted to a thorough discussion of this result. 
In Sec.\ \ref{sec:new}, we first define dephasing and decay channels, then discuss the self-duality of ${\cal L}$, for a single channel in Sec.\ \ref{sec:onegen}\ and for multiple channels in Sec.\ \ref{sec:SDvsH}. We finally analyze how correlated noises give rise to equivalent forms of the master equation in Sec.\ \ref{sec:correlated}.

%%%%%%%%%%%%%%%%%%%%%%%%%%%%%%%%%%%%%%%%%%%%%%%%%%%%%%%%%%%%%%%%

\section{Self-duality and dephasing}
\label{sec:new}

In Sec.~\ref{sec:gentheorem} we have proven our central result that only processes engendered  by a self-dual generator $\mathcal{L}=\mathcal{L}^\sharp$ can be obtained as the average over a classical noise of a unitary evolution, engendered by a  time-dependent self-adjoint Hamiltonian $H_\eta=H_\eta^\dagger$. In this Section we shall investigate in detail the connection between the mathematical concept of self-duality and the physical  notion of decay. By using the findings of Sec.~\ref{sec:gentheorem}, this will enable us to give an answer to the question posed in the title of this article, and explain the laconic abstract.

We will first consider in Sec.~\ref{sec:onegen}  the situation of a single channel, that is a GKLS generator $\mathcal{L}$ with a single term in the operator sum~(\ref{eq:diagl}) and introduce the crucial definition of dephasing and decay channels, that we will need in the following. Then, in Sec.~\ref{sec:SDvsH}, we will move to a generic (multichannel) generator $\mathcal{L}$ and prove that a self-dual generator is the sum of dephasing, \emph{i.e.\ nondecaying}, channels.

\subsection{Single channel} 
\label{sec:onegen}

Let $d$ be the dimension of the Hilbert space. We define a \emph{single channel} as a GKLS generator $\mathcal{L}$ with a single term in the operator sum~(\ref{eq:diagl}), namely
\beq
\mathcal{L} \rho = L \rho L^\dagger - \frac{1}{2} \{ L^\dagger L, \rho \} .
\label{eq:L}
\eeq
We shall say that a single channel $\mathcal{L}$ is \emph{dephasing} if
\beq
\textrm{there exists a basis} \; \{ |e_n\rangle \}_{n=1,\ldots,d} \; \textrm{such that}\;   \mathcal{L} (|e_n\rangle \langle e_n | )=0 \; \textrm{for all} \; n .
\label{eq:dephdef}
\eeq
The basis $\{ |e_n \rangle \}$ is named \emph{stable basis} (under the action of  $\mathcal{L}$). 
This definition is in accord with the general philosophy outlined in Ref.~\cite{BN08}. Observe also that some unitary dynamics are included as limiting cases of the above definition (when the dissipation is null). 

A single channel $\mathcal{L}$ that is not dephasing is called a \emph{decay} channel. Therefore a decay channel admits \emph{no stable basis}: physically, this implies that there are always population flows (except for special initial states). (We notice that these definitions of ``channels" are in line with those adopted in the context of scattering theory.)

We shall prove that 
\begin{equation}
\textrm{$\mathcal{L}$ is a dephasing channel} \iff
\textrm{$L$ is normal},
\label{eq:chaineq}
\end{equation}
that is $[L,L^\dagger]=0$.

The proof goes as follows. Let $\{ |e_n\rangle \}$ be a stable basis. $\mathcal{L}$ dephasing implies that
\beq
L |e_n \rangle \langle e_n | L^\dagger - \frac{1}{2} ( L^\dagger L |e_n \rangle \langle e_n | + |e_n \rangle \langle e_n |L^\dagger L) =0
\label{eq:Lek}
\eeq
for all $n$, and taking expectations over $|e_m \rangle, \; m \neq n$ gives
\begin{equation}
\langle e_m |L |e_n \rangle \langle e_n | L^\dagger  |e_m \rangle =0 , 
\label{eq:Lek1}
\end{equation}
which implies $\langle e_m |L |e_n \rangle =0$. We therefore conclude that $L$ is diagonal in the stable basis, that is
\begin{equation}
L = \sum_{n=1}^d z_n |e_n \rangle \langle e_n | , \quad z_n \in \mathbb{C} 
\label{eq:Lnormal}
\end{equation}
and thus is normal, $[L,L^\dagger] =0$.

Conversely, if $L$ is normal then it can be diagonalized as in~(\ref{eq:Lnormal}), where $\{ |e_n\rangle \}$ is an orthonormal basis. Thus $L$ commutes with $|e_n \rangle \langle e_n |$ for all $n$, and this implies~(\ref{eq:Lek}). This proves the equivalence of the two statements in (\ref{eq:chaineq}).

We now want to establish a link between dephasing (and thus normality of the operator) and self-duality, in the single-channel case.
We shall prove that 
\begin{equation}
 \mathcal{L}=\mathcal{L}^\sharp \quad \Longrightarrow \quad
\textrm{$\mathcal{L}$ is a dephasing channel} 
%\iff L = \sum_{k=1}^n z_k |e_k \rangle \langle e_k | \iff \textrm{$L$ is normal}  ,
\label{eq:chaineq2}
\end{equation}
In words, every single self-dual channel is a dephasing channel. The contrapositive of~(\ref{eq:chaineq2}) reads
\begin{equation}
\textrm{$\mathcal{L}$ is a decay channel} \quad  \Longrightarrow \quad\mathcal{L}\neq\mathcal{L}^\sharp:
\label{eq:chaineqdecay}
\end{equation}
every  decay channel is non-self-dual.

The proof is straightforward. The dual of~(\ref{eq:L}) reads
\beq
\mathcal{L}^\sharp \rho = L^\dagger \rho L - \frac{1}{2} \{ L^\dagger L, \rho \} .
\label{eq:Lsharp}
\eeq
Therefore we get
\begin{equation}
\mathcal{L}\rho =  \mathcal{L}^\sharp \rho   \iff L \rho L^\dagger = L^\dagger \rho L,
\label{eq:Lnormal1}
\end{equation}
and by taking $\rho = \mathbb{I}/d$, we have $[L,L^\dagger]=0$, which in turn, by (\ref{eq:chaineq}), implies~(\ref{eq:chaineq2}).

Two comments are now in order. First, notice that normality of $L$, which is equivalent to a dephasing channel $\mathcal{L}$, does not imply self-duality, and the opposite implication of (\ref{eq:chaineq2}) is not true in general. However, it is possible to give a full characterization of single self-dual channels by strengthening the class of normal operators which act as GKLS operators. Indeed, one gets that
\begin{equation}
\mathcal{L}=\mathcal{L}^\sharp  \iff L^\dagger = \e^{\ii\alpha} L,
\label{eq:chaineq3}
\end{equation}
for some real $\alpha$. That is, a single self-dual channel is characterized by a Hermitian operator modulo a phase: $L= \e^{-\ii\alpha/2} X$, with $X=X^\dagger$. (Remember that a change of phase is a gauge freedom that does not change~$\mathcal{L}$.)
The proof goes as follows. We have proved above that if $\mathcal{L}=\mathcal{L}^\sharp$ then $L$ is normal, and thus is diagonal in some (stable) basis as in~(\ref{eq:Lnormal}). By plugging~(\ref{eq:Lnormal}) into the right hand side of~(\ref{eq:Lnormal1}), we get
\begin{equation}
z_{m} z_{n}^* = z_{m}^* z_{n},
\end{equation}
for all $m,n$, which implies that $z_n = \e^{-\ii \alpha/2} x_n$, with $x_n$ real for all $n$ and for some real ($n$-independent) phase $\alpha$. The converse is immediate. 
The generalization of this characterization for dilations that include also Poisson processes has been proved by K\"ummerer and Maassen in \cite{KM87}, where the detailed balance condition \cite{Kos1} plays a significant role.

Second, the extension of the above results to a multichannel  process~(\ref{eq:diagl}) has to confront the noncommutativity of the GKLS operators $L_k$. Suffice it to say that, in general, in the presence of several dephasing channels there exists \emph{no} stable basis. Indeed, each single channel admits a stable basis, but two  bases are incompatible if the corresponding GKLS operators do not commute. Therefore the net effect of several dephasing channels could be mistaken with decay by a naive application of  definition~(\ref{eq:dephdef}). We will deal with such a general situation in the next subsection.
%Sec.~\ref{sec:SDvsH}.

\subsection{Multiple channels} 
\label{sec:SDvsH}

In the previous subsection we have proved that a single self-dual channel is a dephasing channel, by using the characterization of a dephasing channel in terms of the normality of its GKLS operator. 
Now we shall establish such a link in the general case of a multichannel process. 
We shall prove that 
\begin{equation}
 \mathcal{L}=\mathcal{L}^\sharp \quad \Longrightarrow \quad \mathcal{L} = \sum_{k=1}^N \mathcal{L}_k
\quad \textrm{with $\mathcal{L}_k$ a  dephasing channel}.
\label{eq:chaineqmulti}
\end{equation}
In words, every  self-dual process is composed by one or more dephasing channels. The contrapositive of~(\ref{eq:chaineqmulti}) states that if a process cannot be decomposed into a sum of dephasing channels, then its generator is non-self-dual:  the presence of a \textit{bona fide} decay is symptomized by  a non-self-dual $\mathcal{L}$.

We will get~(\ref{eq:chaineqmulti}) by proving the stronger result that  $\mathcal{L}=\mathcal{L}^\sharp$ implies 
each channel $\mathcal{L}_k$ to have a Hermitian GKLS operator $L_k$ and thus by~(\ref{eq:chaineq3}) to be self-dual:
\begin{equation}
\mathcal{L}_k = \mathcal{L}_k^\sharp, \qquad \textrm{for all } k.
\end{equation}
This in turn, by~(\ref{eq:chaineq2}), implies~(\ref{eq:chaineqmulti}).

Here is the proof. Consider a (multichannel) dissipative $\mathcal{L}$ as in~(\ref{eq:diagl}) and its dual~(\ref{eq:diagldual}).
The self-duality condition, $\mathcal{L}\rho=\mathcal{L}^\sharp\rho$,
requires that the following relation be satisfied 
\begin{equation}
\sum_k L_k\rho L_k^\dagger=\sum_k L_k^\dagger\rho L_k
\label{eq:selfdual}
\end{equation}
for any $\rho$. 
Notice that (\ref{eq:selfdual}) implies the equality
\begin{equation}
\sum_k L_kL_k^\dagger=\sum_k L_k^\dagger L_k
\end{equation}
and therefore the dual of $\mathcal{L}$ reads 
\begin{equation}
\mathcal{L}^\sharp\rho=-\frac{1}{2}\sum_k \Bigl(\{L_k^\dagger L_k,\rho\} - 2 L_k^\dagger\rho L_k\Bigr)
= -\frac{1}{2}\sum_k \Bigl(\{L_k  L_k^\dagger,\rho\} - 2 L_k^\dagger\rho L_k\Bigr).
\end{equation}
We can write any self-dual $\mathcal{L}$ as
\begin{equation}
\mathcal{L}\rho={1\over2}(\mathcal{L}+\mathcal{L}^\sharp)\rho,
\end{equation}
which implies that every GKLS operator appears in pair with its Hermitian conjugate.
Explicitly, we have
\begin{eqnarray}
\mathcal{L}\rho
&=-{1\over4}\sum_k\Bigl( \{L_k^\dagger L_k,\rho\} +\{L_kL_k^\dagger,\rho\} -2 L_k\rho L_k^\dagger -2 L_k^\dagger\rho L_k
\Bigr)\nonumber\\
&=-{1\over2} \sum_k\Bigl(\{X_k^2,\rho\} -2 X_k\rho X_k  + \{Y_k^2,\rho\}  -2Y_k\rho Y_k\Bigr),
\label{eq:LHerm}
\end{eqnarray}
where the Hermitian operators $X_k$ and $Y_k$ are defined by $L_k=X_k+\ii Y_k$.
The last expression means that a self-dual $\mathcal{L}$ can always be described by the sum of Hermitian GKLS operators.
This completes the proof. 

It is worth noticing that the decomposition of a generator $\mathcal{L}$ in terms of single channels is not unique, as for example is manifestly shown in the first and the second line of~(\ref{eq:LHerm}). Thus an evolution could be built up by a sum of decaying channels whose net effect is nevertheless self-dual, and thus, by~(\ref{eq:chaineqmulti}), equivalent to a sum of purely dephasing channels. A paradigmatic example is  the two-level atom in a thermal photon bath considered in~(\ref{decay2}): when the temperature goes to infinity it happens that $\gamma' (1+n) \sim \gamma' n$, and the net population transfer between the two atomic levels goes to zero. This again is related to  a detailed balance condition~\cite{Kos1}.

Conclusions (\ref{eq:centrth}) and (\ref{eq:chaineqmulti}) are the central results of this article. 
As stressed before, the contrapositive of~(\ref{eq:chaineqmulti}) states that
decay can only be obtained by a non-self-dual $\mathcal{L}$. Therefore an interpretation involving ``classical noise," in the sense of Eqs.~(\ref{eq:ham})--(\ref{eq:HermHS}) and (\ref{eq:centrth}), is untenable.

\section{Correlated noises and equivalent forms of the master equation}
\label{sec:correlated}
We elaborate here on equivalent forms of the master equation and their corresponding stochastic Schr\"odinger equations.
So far, our analysis has focused on noise terms of the type
\begin{equation}
\sum_{k=1}^N L_k \d W_k, 
\label{noiseterms}
\end{equation}
with $N\leq d^2-1$, $d$ being the dimension of the system, with generally non-Hermitian operators $L_k$ and \emph{real} independent noises $\d W_k$ such that 
\begin{equation}
\langle \d W_k \d W_l \rangle= \delta_{kl} \d t.
\label{ww}
\end{equation}
This ansatz yields a master equation with a diagonalized Kossakowski matrix, as in Eq.~(\ref{me}).

However, this is clearly not the only option. For example, one can decide to work with Hermitian GKLS operators
and expand the $L_k$'s in terms of $d^2-1$ linearly independent \emph{Hermitian} operators $\lambda_j$  [e.g., su$(d)$ operators]
\begin{equation}
L_k = \sum_{j=1}^{d^2-1} c_{kj} \lambda_j ,
\label{hermgen}
\end{equation}
where $c_{kj}$ are the complex coefficients of the expansions.
In such a case one ends up with $d^2 -1$ \emph{complex} noise terms 
\begin{equation}
\sum_{j=1}^{d^2-1} \lambda_j \d Z_j, \qquad \d Z_j=\sum_{k=1}^N c_{kj} \d W_k,
\label{cnoise}
\end{equation}
that are in general not independent:
\begin{equation}
\langle \d Z_i^* \d Z_j \rangle = a_{ij} \d t, \qquad \langle \d Z_i \d Z_j \rangle = b_{ij} \d t,
\label{cnoisedep}
\end{equation}
with 
\begin{equation}
a_{i j}= \sum_{k=1}^N c_{ki}^* c_{kj}, \qquad b_{ij} =  \sum_{k=1}^N c_{ki} c_{kj}.
\label{abc}
\end{equation}

The covariance matrix $a$ is positive semi-definite, $a = a^\dagger$ and $a \ge0$, while the ``relation matrix" $b$ is symmetric, $b = b^T$, satisfying Picinbono's condition $a^*-b^\dag a^{-1}b\ge0$ (with the inverse $a^{-1}$ defined on the support of $a$) \cite{Picinbono}. Such conditions guarantee the positivity of the complex noise matrix.
The dissipative part of the corresponding master equation reads
\begin{equation}
\mathcal{L} \varrho  =
-\frac{1}{2}\sum_{i,j} a_{ij} ( \{ \lambda_i \lambda_j, \rho \} - 2 \lambda_j \rho \lambda_i ),
\label{newME}
\end{equation}
instead of~(\ref{me}).
Observe that the noise correlations yield the Kossakowski matrix $a_{ij}$. 

In order to obtain the master equation~(\ref{newME}) from a stochastic Schr\"odinger equation with the complex noise terms~(\ref{cnoise}),
the first condition in Eq.~(\ref{cnoisedep}), $\langle \d Z_i^* \d Z_j \rangle = a_{ij} \d t$, is crucial, while the second one, $\langle \d Z_i \d Z_j \rangle= b_{ij} \d t$, is not needed and the relation matrix $b$ can be arbitrary, as long as $b$ satisfies Picinbono's condition.
However, in order to go from the stochastic Schr\"odinger equation with the complex noises $\d Z_i$ in Eq.~(\ref{cnoise}) to the one~(\ref{eqn:SSEL}) with the real independent noises $\d W_i$, by diagonalizing the covariance matrix $a$, the relation matrix $b$ should be appropriately chosen in order to get the minimal number of real noises. 
Note that there are $2d^2-2$ real noises (real and imaginary parts) in the $d^2-1$ complex noises $\d Z_i$, but only $d^2-1$ real noises $\d W_i$ suffice for the stochastic Schr\"odinger equation~(\ref{eqn:SSEL}), with the rest of the degrees of freedom being redundant. 
The right choice of $b$ is the following.
We diagonalize $a$ as $a_{ij}=\sum_k\gamma_k U_{ki}^*U_{kj}$, with a unitary matrix $U$ and positive semi-definite eigenvalues $\gamma_i$.
Then, we choose the relation matrix $b$ as $b_{ij}=\sum_k\gamma_k U_{ki}U_{kj}$, which makes half of the real noises in $\d Z_i$ irrelevant (vanishing).
On the other hand, while the choice of the relation matrix $b$ does not affect the master equation~(\ref{newME}), it does affect the Stratonovich Hamiltonian.

%%%%%%%%%%%%%%%%%%%%%%%%%%%%%%%%%%%%%%%%%%%%%%%%%%%%%%%%%%%%%%%%

\section{Answer(s).}
\label{sec-concl}

Let us summarize the overall picture of our results.
For a generic GKLS generator $\mathcal{L}$ of a master equation (\ref{me}) with multiple channels, by combining the implications (\ref{eq:centrth}) and (\ref{eq:chaineqmulti}), we get
\begin{eqnarray}
&&\hspace*{-6truemm}
H_\eta=H^\dagger_\eta\iff H_2^S=0
\iff L_k=L_k^\dagger,   \;  \forall k
\nonumber\\
&&\hspace*{-6truemm}
\hphantom{H_\eta=H^\dagger_\eta\iff H_2^S=0}\ \,%
\Longrightarrow\ \mathcal{L}=\mathcal{L}^\sharp
\nonumber\\
&&\hspace*{-6truemm}
\hphantom{H_\eta=H^\dagger_\eta\iff H_2^S=0}\ \,%
\Longrightarrow\ \mathcal{L} = \sum_{k=1}^N \mathcal{L}_k
\ \textrm{with $\mathcal{L}_k$ a  dephasing channel}.
\nonumber\\
\label{eq:centrth11}
\end{eqnarray}
For each single channel $\mathcal{L}_k$, by combining the implications (\ref{eq:chaineq}), (\ref{eq:chaineq2}), and (\ref{eq:chaineq3}), we have
\begin{eqnarray}
&&
\mathcal{L}_k=\mathcal{L}_k^\sharp
 \iff L_k^\dagger = \e^{\ii\alpha_k} L_k
\nonumber\\
&&\hphantom{\mathcal{L}_k=\mathcal{L}_k^\sharp}
 \ \,\Longrightarrow\ %
[L_k,L_k^\dagger]=0 \iff
\textrm{$\mathcal{L}_k$ is a dephasing channel}.
%\label{eq:chaineq}
\end{eqnarray}

The answer to the question posed in the title of this article is negative: decay cannot be ascribed to a ``classical" noise process, where the connotation of the term ``classical" has to be understood according to 

Eqs.~(\ref{eq:LLdag})--(\ref{eq:iHam}), or in general Eqs.\ (\ref{eq:ham})--(\ref{eq:HermHS}) and (\ref{eq:centrth}), without the \emph{caveat} of taking the average over the noise, as e.g.\ in Eq.~(\ref{eq:probOK}).

This is a consequence of the chain of equivalence (\ref{eq:centrth11}), that is valid for master equations with an arbitrary finite number of multidimensional GKLS operators. Dephasing is inextricably related to self-dual generators and as a consequence the opposite process, decay, can only be ascribed to non-self-dual maps.  
Physical interpretations involving ``classical noises" only apply to the former process.
On the contrary, the latter process entails non-Hermitian Hamiltonians (and imaginary optical potentials \emph{\`a la} Fermi \cite{fermizinn}): probability would no longer be conserved. 

There is, however, a second possible answer to our question: yes, decay can be ascribed to a ``classical" noise process, if we relax the condition~(\ref{eq:probOK2}) of probability conservation in individual realizations, and just require probability conservation on average:
During the stochastic process, sometimes particles are absorbed by the environment, sometimes they are released, with a null average net flux. This is what we called \emph{weak} conservation of probability before Eq.~(\ref{eq:probOK}).

\section*{Acknowledgments}
We thank John Gough, Burkhard K\"ummerer and Hans Maassen for their important suggestions and an interesting email exchange.
DB, PF, GG and SP would like to thank the Department of Physics and the Department of Applied Physics of Waseda University for the hospitality within the Top Global University Project from the Ministry of Education, Culture, Sports, Science and Technology (MEXT), Japan.
This work was partially supported by INFN through the project ``QUANTUM", and by the Italian National Group of Mathematical Physics (GNFM-INdAM).
DB acknowledges support from the EPSRC Grant No.\ EP/M01634X/1.
KY was supported by the Grant-in-Aid for ScientificResearch (C) (No.\ 26400406) from the Japan Society for the Promotion of Science (JSPS) and by the Waseda University Grant for Special Research Projects (No.\ 2016K-215).
HN is supported by the Waseda University Grant for Special Research Projects (No.\ 2016B-173).

\end{document}